\documentclass[]{emulateapj}

\usepackage{natbib}
\usepackage{graphicx}
\usepackage{rotating}
\usepackage{epsfig}
\usepackage{epstopdf}
\usepackage{enumitem}

\begin{document}

\shorttitle{Kinetic turbulence in the fast solar wind }
\shortauthors{Roberts, Li, and Li}

\title{Kinetic plasma turbulence in the fast solar
wind measured by Cluster} 


\author{ O.W. Roberts$^1$,  X. Li$^1$, and B. Li$^2$,
}  \affil{1. Institute of Mathematics and Physics, Aberystwyth
University, United Kingdom SY23 3BZ, \email{xxl@aber.ac.uk}. \\
2. School of Space Science and Physics,
Shandong University, Weihai 246209, China.\\
}




\begin{abstract}
The k-filtering technique and wave polarization analysis are applied
to Cluster magnetic field data to study  plasma turbulence at the
scale of the ion gyroradius in the fast solar wind.
Waves are found propagating in directions nearly perpendicular to
the background magnetic field at such scales. The frequencies of
these waves in the solar wind frame are much smaller than the proton
gyro-frequency. After the wave vector ${\bf k}$ is determined at
each spacecraft frequency $f_{sc}$, wave polarization property is
analyzed in the plane perpendicular to ${\bf k}$. Magnetic
fluctuations have
 $\delta B_\perp>\delta B_\parallel$ (here the $\parallel$ and
$\perp$ refer to the background magnetic field ${\bf B}_0$). The
wave magnetic field has right-handed polarization at propagation
angles $\theta_{\bf kB}<90^\circ$ and $>90^\circ$. The magnetic
field in the plane perpendicular to ${\bf B}_0$ however has no clear
sense of a dominant polarization but local rotations. We discuss the
merits and limitations of linear kinetic Aflv\'en waves (KAWs) and
coherent Alfv\'en vortices in the interpretation of the data. We
suggest that the fast solar wind turbulence may be populated with
KAWs, small scale current sheets and Alfv\'en vortices at ion
kinetic scales.


\end{abstract}

\maketitle


\section{Introduction}

The solar wind is a natural laboratory to investigate plasma
turbulence. It is well known that in the inertial range, at which
the usual magnetohydrodynamic (MHD) description is still valid,
magnetic turbulence is strongly anisotropic: for a given wave number
${\bf k}$, magnetic fluctuation energy is much more concentrated at
quasi-perpendicular propagation ($k_\perp \gg k_\parallel$) than it
is at quasi-parallel propagation ($k_\perp \ll k_\parallel$ )
\citep{Shebalin, Goldreich95, matthaeus90, bieber96, Horbury05,
Dasso05}. Numerous measurements find the Kolmogorov $k^{-5/3}$
spectrum of magnetic field fluctuations in the inertial range and a
steeper spectrum at ion kinetic scales (which is often called the
dissipation range where MHD description breaks down).
A spectral break point around $k\rho_i \approx 1$ (where $\rho_i$ is
the ion thermal gyroradius) or $k d_i \approx1$ (where $d_i$ is the
ion inertial length), which marks the end of the $k^{-5/3}$ inertial
range, suggests possible initiation of kinetic dissipation processes
at ion scales while turbulent cascade continues to operate at the
same scales  and at smaller scales, up to electron gyroradius
(Alexandrova et al. 2009, 2012).
It is an open question exactly which scale is responsible for the
spectral break (see a recent discussion on this topic by
\cite{Bour2012}). A view to account for the observed spectral
steepening at high frequencies (ion scales) is to interpret the
spectral steepening as evidence of kinetic Alfv\'en waves
\citep{Leamon98, Bale05, Howes08, Schekochihin09, Howes10,
Sahraoui10, Salem2012}, or whistler waves \citep{Biskamp96,
Lietal2001, Stawicki,
 Garysmith09} under the assumption that although linear waves are
unable to produce nonlinear cascade, they may still approximately
describe the nature of turbulence at ion kinetic scales.
An alternative view is that 2D structures (such as current sheets,
coherent magnetic vortices) populate the fluctuations at these
scales and  have been observed in the ionosphere, magnetosphere and
magnetosheath (Chmyrev et al. 1988, Volwek et al. 1996, Sundkvist et
al. 2005, Alexandrova et al. 2006).



Using magnetic field data recorded simultaneously by the four
Cluster spacecraft and assuming that turbulence contains many
structures on scales to be measured and the time series are at least
weakly stationary \citep{Pincon}, the k-filtering technique assumes
plane wave geometry and has been applied to the magnetosphere and
magnetic reconnection \citep{Sahraoui04, Grison05, 
Narita05, Eastwood09, Huang10}. It is well-known that the
k-filtering method is subject to a spatial aliasing effect
\citep{Pincon}. Great care must be taken to eliminate or minimize
the spatial aliasing. This can be realized by setting the maximum
wave number and spacecraft frequency to be analyzed properly
\citep{Sahraoui10jgr}. Its application to the solar wind turbulence
is limited and results are inconclusive: Sahraoui et al. (2010a)
found that KAWs populate in the solar wind turbulence ion scales
while Narita et al. (2011) concluded that linear Vlasov theory is
insufficient to describe the plasma turbulence and turbulent cascade
is at work. It should be noted that the data studied in
\citet{Sahraoui10} were taken during a coronal mass ejection
\citep{jian}.  Narita et al. (2011) used data when the tetrahedral
configuration of the Cluster spacecraft was not optimal: the
planarity $P$ and elongation $E$, which describe the degree that the
four Cluster spacecraft are close to perfect tetrahedron (Robert et
al. 1998), were such that $ P>0.3$ and $E \geqslant 0.1$,
undesirable to apply the k-filtering (Sahraoui et al. 2010b) in such
geometries.

In this paper, we present a new study of Cluster data to study solar
wind plasma turbulence at ion kinetic scales
 by combining the
k-filtering technique and wave polarization analysis. Although
unable to determine wave propagation direction, polarization
analysis supports the interpretation of KAWs in the turbulence
dissipation range when interplanetary magnetic field is in the
direction nearly perpendicular to the solar wind (He et al. 2012).
We present data analysis in section 3 and discussions on the
interpretation of the data in section 4. We summarize our findings
and conclude the paper in section 5.


\section{Data}

Table 1 summarizes key parameters of four periods (P1, P2, P3 and
P4) on 31/01/2004 and 29/02/2004 when the Cluster spacecraft were in
the ambient fast solar wind. The mean parameters of the periods are:
$B_0$ the strength of the averaged magnetic field, $n$ the total ion
density, $\beta$ the ion plasma beta (ratio between ion parallel
thermal pressure and magnetic field pressure), $V_f$ the solar wind
speed, $f_{ci}=eB_0/(2\pi m_p)$ proton gyrofrequency, $v_A$ the
Alfv\'en speed, $E$ elongation, $P$ planarity, $\theta_{VB}$ the
angle between the solar wind and ${\bf B}_0$, $T_\perp/T_\parallel$
hot ion temperature anisotropy, $T_e/T_i$ the ratio of electron to
ion temperature, $\rho_i$ ion thermal gyro-radius,
$d_i=v_A/\Omega_p$ ion inertial length, and $n_\alpha/n_p$ the
abundance of alpha particles (fully ionized helium).  The $n_p$ and
$n_\alpha$ are the densities of protons and alpha particles.
During the chosen intervals, the magnitude of the magnetic field was
quite stable and there were no obvious discontinuities (see the raw
magnetic field data from C4 in P3 in Fig. 1a). Both planarity $P$
and elongation $E$ are smaller than 0.1 during the
periods. 

The magnetic field data were from the Fluxgate Magnetometer (FGM)
\citep{Balogh01}. FGM measures components of the magnetic field in
the GSE (geocentric solar ecliptic) coordinate system. In the
coordinate system, positive $x$ points from the Earth to the Sun,
and positive $z$ points to the ecliptic north pole. We use full
resolution magnetic field data (22 samples/sec). The average
distance $d$ between the spacecraft was $d \sim 200$km in the four
periods. The magnetic field was primarily oriented in the direction
perpendicular to the solar wind direction so direct magnetic
connection with the bow shock does not exist. On 31 Jan., the ion
plasma data from the Hot Ion Analyzer (HIA) instrument
\citep{reme2001} (with spin resolution) are available from C1
spacecraft, and on 29 Feb. they are available from both C1 and C3
(the difference between them is very small).
Electron temperature data 
are obtained from the Plasma Electron and Current Experiment (PEACE)
instrument (Johnstone, 1997) onboard the C4.

\section{Data analysis}


\subsection{K-filtering}

Fig. 1b shows the Fourier power spectra of the three magnetic field
components of data from C4 during the periods P3 and P4. The spectra
are typical of the turbulent magnetic field fluctuations in the
solar wind. At relatively low frequencies (0.007-0.4Hz) the
fluctuations have an $f^{-5/3}_{sc}$ Kolgomorov power law. At a
breakpoint $f_{sc} \sim 0.4-0.5$Hz, the spectra steepen with a
spectral index of about -3.5. The spectra become flattened again at
the second breakpoint roughly at 2.4Hz due to FGM reaching the noise
floor \citep{Balogh01}.

The k-filtering method is a measurement technique designed for
multipoint measurements which does not require Taylor's frozen-in
flow hypothesis (Taylor 1938): using plane wave assumption, it
estimates the spectral energy density $P(\omega, {\bf k})$ in
Fourier space (angular frequency $\omega $ and wave vector ${\bf k}$
domains) by combining several time series recorded simultaneously at
different locations in space. The k-filtering method uses a filter
bank approach \citep{pincon1998,Tjulin2005} by adopting the random
phase approximation. The filter is dependent on $\omega$ and ${\bf
k}$, and is designed in such a way that it absorbs all wave field
energy except those plane waves with $\omega$ and ${\bf k}$.

Similar to temporal Fourier analysis, if the spacecraft distance is
$d$, the maximum wave number the spacecraft can measure is
$k_{max}=\pi /d$ \citep{pincon1998,Sahraoui10jgr}. Due to the use of
Fourier analysis, spatial aliasing will occur when the spacecraft
configuration does not distinguish two plane waves differing only in
 wave vectors by $\Delta {\bf k}$:
\begin{equation}
\Delta {\bf k} \cdot {\bf r}_{ij}=2 \pi n_{ij}, ~~~~1\leq i < j \leq
N
\end{equation}
where ${\bf r}_{ij}={\bf r}_i-{\bf r}_j$ is the separation vector
between two spacecraft ($i$ and $j$), $n_{ij}$ is an integer, and
$N$ is the number of spacecraft. For Cluster mission ($N$=4), the
solution to the above equation, can be written as (Neubauer \&
Glassmeier 1990, Tjulin et al. 2005):
\begin{equation}
\Delta {\bf k}=n_{14}\Delta {\bf k}_1 +n_{24} \Delta {\bf k}_2 +
n_{34} \Delta {\bf k}_3,
\end{equation}
where
\begin{equation}
\begin{array}{l}
\Delta {\bf k}_1 =2 \pi {\bf r}_{24}\times {\bf r}_{34}/V, \\
\Delta {\bf k}_2 =2 \pi {\bf r}_{34}\times {\bf r}_{14}/V, \\
\Delta {\bf k}_3 =2 \pi {\bf r}_{14}\times {\bf r}_{24}/V,
\end{array}
\end{equation}
and
$$
V={\bf r}_{14} \cdot ({\bf r}_{24} \times {\bf r}_{34})
$$

In our analysis, only wave energy peaks in the ${\bf k}$ space
centered at ${\bf k}=0$ are counted by assuming they are due to
waves physically present in the solar wind (not due to aliasing).
This ${\bf k}$ space is given by
\begin{equation}
{\bf k} =\mu_1 \Delta {\bf k}+1 +\mu_2 \Delta {\bf k}_2 +\mu_3
\Delta {\bf k}_3, ~~~\mbox{where }~
 -1/2\leq \mu_{1,2,3} \leq 1/2.
\end{equation}
A wave with a wave vector in this region will not produce any
aliased energy peak in the region. Outside this region, wave energy
peaks will be dropped from our analysis. Obviously, a wave with a
wave vector outside this region may also produce aliased energy
peaks inside the region. However, \emph{this issue is not expected
to influence our analysis in a significant way due to two reasons}.
Firstly, we may generally assume that turbulence at smaller wave
numbers contains more power than at larger wave numbers. Hence, the
aliased energy peaks produced by larger wave numbers may be too weak
to be noticeable when the wave energy peaks of small wave numbers
are present. Secondly, turbulence with larger wave numbers may also
have higher frequencies (at least for normal plasma modes). As a
result, the power of waves with larger wave numbers may be filtered
when we analyze the power of waves at low frequencies. We notice
that when the four Cluster spacecraft form a regular tetrahedron
configuration, the magnitude of the three wave vectors in Eq. (3) is
greater than $2k_{max}$. Therefore aliased energy peaks have
substantial difference in their wave numbers. This fact is strongly
in favor of our first argument since it is generally known that the
solar wind turbulence power rapidly drops with increasing wave
numbers.

The two vertical dashed lines in Fig. 1 represent the minimum and
maximum frequencies $f_{min}=0.07$Hz and $f_{max}$=1.1Hz between
which the k-filtering technique is applied in this paper. To avoid
or minimize the spatial aliasing, a maximum spacecraft frequency
$f_{max}$ has to be set corresponding to the maximum wave number
$k_{max}$ \citep{Sahraoui10jgr}. This is also necessary to avoid a
frequency aliasing effect (Narita et al. 2010). Note, it is
fortunate that generally high spacecraft frequency corresponds to
large $k$. In the solar wind rest frame, the maximum frequency is
$k_{max} v_{ph}/(2\pi)$, where $v_{ph}=\mbox{Max}(v_A,c_s)$ and
$c_s$ is the ion sound speed. The choice of $v_{ph}$ is equivalent
to the assumption that there are no whistler waves at scales near
$k_{max}$ in the solar wind. This will be verified later on for the
solar wind data we analyzed. If whistler waves do exist, the choice
of $v_{ph}$ and the maximum frequency must be dealt with
accordingly.
 Since the solar wind is supersonic and
super Alfv\'enic, the maximum frequency $f_{max}$ may be set at
$k_{max} (V_f-v_{ph})/(2\pi)=1.32$Hz due to Doppler effect (here
$V_f$ is the solar wind speed). Note, it is likely that the wave
vectors will deviate from the solar wind direction at an angle
$\theta_{kV_f}$. In such a case, a spacecraft frequency higher than
$f_{max} = k_{max} (V_f-v_{ph})/(2\pi) \cos \theta_{kV_f}$ will
correspond to a wave number larger than $k_{max}$. In the periods we
studied, it is found that the vectors can deviate up to
$\theta_{kV_f}=30^\circ$ from the solar wind direction at the
highest frequencies (and wave numbers). In this work we set
$f_{max}=1.1$Hz. A key reason of choosing this maximum frequency is
that above this frequency the noise level of FGM is generally
believed to be high (according to FGM PI Elizabeth Lucek, private
communication). In fact, within $1.1<f_{sc}<1.32$Hz, we are still
able to use the k-filtering if the FGM noise is low. We chose 1.1Hz
as the upper limit to avoid producing unphysical results due to the
FGM noises.

The $f_{min}$ value is fixed by choosing $k_{min}=k_{max}/25$ and
 $f_{min} \approx k_{min}V_f/(2\pi)$ so
that the wave vectors are computed with relative accuracy better
than 10\% (1\% at the highest frequency) \citep{Sahraoui10jgr}.
Obviously, the minimum frequency is also limited by the number of
sampling points available in a dataset. It is important to point out
a limit of k-filtering technique in determining the solar wind
turbulence power. For wave vectors almost perpendicular to the solar
wind direction, the wave number component parallel to the solar wind
flow is $k_\parallel \approx 0$. In this case, if the wave has a
small frequency $\omega_{plas}$ in the solar wind frame, the Doppler
shifted frequency $f_{sc}=(\omega_{plas}+V_fk_\parallel)/(2\pi)$ may
be lower than $f_{min}$. Such a wave will not be resolved by
k-filtering. However, as long as the wave number is in the wave
number space described by Eq. (4) and the Doppler shifted frequency
$f_{sc}$ is greater than $f_{min}$, the wave will be resolved by the
k-filtering.

By scanning the ${\bf k}$ space, the k-filtering technique is used
to determine the strongest wave power $P(\omega_{sc},~{\bf k}$) and
the corresponding wave vector ${\bf k}$ at each $f_{sc}$. The wave
power in the solar wind frame $P(\omega_{plas},~{\bf k})$ is then
determined using the Doppler shift $\omega_{plas}=\omega_{sc}-{\bf
k} \cdot {\bf V}_f,$ and the wave dispersion relation $\omega_{plas}
=\omega_{plas}({\bf k})$ is obtained. Four studied intervals P1, P2,
P3 and P4 are shown respectively in black, blue, green and red in
Fig. 2. CIS onboard moments are used for ion parameters. A small
correction is made to the solar wind speed since a few percent of
the ions are minor ions (mainly fully ionized helium) and the CIS
onboard moments are calculated by assuming all the detected ions are
protons (HIA/CIS measures the ion energy per charge
\citet{reme2001}). The abundance of helium ions can be found from
the ion velocity distribution function (VDF) measurements by
assuming that protons and helium ions have the same flow speed $V_s$
so two populations can be separated (Marsch et al. 1982). Given the
solar wind speed from the CIS onboard moment data $V_f$, one finds
\begin{equation}
V_s=\sqrt{1+n_\alpha/n_p \over 1+2n_\alpha/n_p}V_f
\end{equation}
 We use $V_s$ (instead of $V_f$) to compute $\omega_{plas}$.

\begin{table}[b]
\caption{\label{tab:table1}%
Average plasma parameters during chosen intervals that
k-filtering technique is applied (data are from CIS, FGM and PEACE).
}
\begin{tabular}{lcccc}
& Jan. 31 (P1) &Jan. 31 (P2) & Feb. 29 (P3) & Feb. 29 (P4)
\\
& 14:30-14:40& 14:45-14:55&04:10-04:20&04:25-04:35\\
B (nT) &  8.45 &7.97 &9.56&9.34\\
n(cm$^{-3}$) &   3.47 & 3.25 & 2.88 & 2.73\\
$\beta$&  0.62 & 0.72& 0.73 & 0.67 \\
$V_f$ & 613 & 609 & 646 & 657 \\
$f_{ci}$ & 0.129 & 0.122 & 0.146 &0.142\\
$v_A$ & 99.1 & 96.2 & 123.1 & 123.4\\
E & 0.05 &0.04 &0.01 &0.02 \\
P & 0.07 &0.06 &0.03 &0.01 \\
$\theta_{VB}$ & 75.1$^\circ$ & 66.6$^\circ$& 78.6$^\circ$ &
84.1$^\circ$ \\
$T_{i\perp}/T_{i\parallel}$ &1.41&1.28&1.26&1.46 \\
$T_{e}/T_i$ &N/A & N/A& 0.37 & 0.39 \\
$\rho_i$ (km) &   115&121&129&137\\
$d_i$ (km) & 122&126& 134 & 138 \\
 $n_\alpha/n_p$ & 1.4\% & 1.3\% & 0.38\% & 0.2\%
\end{tabular}
\end{table}

As shown in Fig. 2a, wave vectors are mainly in directions
quasi-perpendicular to ${\bf B}_0$ with $\langle \theta_{\bf
kB}\rangle= 81\sim 90.1^\circ$, similar to previous work
\citep{Sahraoui10}. The wave vectors and the solar wind flow make
moderate angles, $\langle\theta_{\bf
kV_f}\rangle=13-30^\circ$, 
so generally waves are propagating in directions not far away from
the solar wind direction. The error bars (Fig. 2a) at low
frequencies are significantly larger than at higher frequencies,
reflecting larger relative uncertainty when determining smaller wave
numbers \citep{Sahraoui10jgr}.

Fig. 2b displays the measured dispersion relation (filled dots). The
error bars in the figure mainly come from an assumed 3.5\%
uncertainty in the solar wind flow speed in addition to the
uncertainty in the wave vector \citep{Sahraoui10jgr}.  At the energy
channel (2359.28eV) of HIA/CIS where the peak particle flux of the
solar wind is measured during the studied periods, the energy
resolution is about $\leq 7$\%. Hence, the error from a 3.5\%
uncertainty in the solar wind speed at 650km/s with Alfv\'en speed
at $v_A=120$km/s and $\theta_{kVf}=20^\circ$ and $kv_A/\Omega_p
=1.4$ can be estimated as
$$
{\Delta \omega_{plas} \over \Omega_p} \sim { kv_A \over \Omega_p}
\times {0.035V_f \over v_A} \cos 20^\circ \sim 0.25.
$$
However error bars in the dispersion plot of Sahraoui et al. (2010)
are puzzlingly small.
 Plotted in Fig. 2b are the dispersion relations of waves
propagating in some measured propagation angles, $\theta_{{\bf
kB}}$. They include fast and Bernstein waves propagating at
89.5$^\circ$ (red solid and dot-dashed), and KAWs propagating at
80$^\circ$, 85$^\circ$ and 89$^\circ$. At very high $k$ at which our
data are unable to cope, the branches of KAWs are highly dispersive
and are named ``oblique whistler" waves (Sahraoui et al. 2012). In
computing the dispersion, the abundance of alpha particles is 2\%
and the alphas and protons have the same thermal speed.


 Note that  $\omega_{plas}/\Omega_p$ can be
negative. It is found that 95 data points of $\omega_{plas}$ are
positive while 36 are negative.  Most of the negative frequencies
are small and 27 of the negative data points are within the
uncertainties of small positive frequencies. The largest
uncertainties of $\omega_{plas}/\Omega_p$ in Fig. 2b are $\pm 0.25$
at large $kv_A/\Omega_p \approx 1.4$. The uncertainties are about
$\pm 0.04 $ when $kv_A/\Omega_p=0.2$. We found that 9 of the 36
negative frequencies may have to be interpreted as waves propagating
in the sunward direction in the solar wind frame. Statistically they
are less important and we will defer their investigation to  a
future study.


In Fig. 2d, power spectral density of magnetic field fluctuations as
a function of $k_y$ at spacecraft-frame frequency 0.51Hz in P3, a
well-behaved peak, is shown.  In the $x$ and $z$ directions, the
peaks are narrower than in the $y$ direction. Therefore, the
dominant $k$ is well defined in the data. Fig. 2e displays the
measured magnetic field $k_{\perp}$ spectra of the four intervals.
Few data points exist at small $k$. Data from the four intervals are
combined. The two solid lines show two power laws with spectral
indices of -5/3 and -3.5. The spectrum roughly reveals two power
laws\citep{Sahraoui10}: a Kolmogorov scaling $\sim k^{-5/3}_\perp$
at smaller $k_\perp$ above a breakpoint at $k_\perp \rho_i \approx
0.4-0.5$. The spectrum steepens to a $k^{-3.5}_\perp$ scaling in an
ion dissipation range $k_\perp \rho_i \in [0.5-1.5]$.


\subsection{Polarization analysis in the plane perpendicular to
${\bf k}$}

Once ${\bf k}=(k_x,k_y,k_z)$ is found, a primed Cartesian coordinate
system is constructed to study wave polarization. The direction of
${\bf k}$ is along the $z'-$axis with a unit vector ${\bf
e}_{z'}={\bf k}/k$. The unit vector along the primed $x'-$axis is
${\bf e}_{x'}$, and ${\bf e}_{z'}={\bf e}_{x'}\times {\bf e}_{y'}$.
Let
\begin{equation}
{\bf e}_{x'}=[-k_yA/k_x,A,0]
\end{equation}
describe the three components of ${\bf e}_{x'}$ in the GSE
coordinates, where $A={k_x \over \sqrt{k^2_x+k^2_y}}$.  Then the
three components of magnetic field fluctuations in the GSE
coordinates are projected on the primed coordinates.
A Morlet wavelet transform, a natural bandpass filter, is used
 (He et al. 2012) and the time series
reconstructed at a frequency $f_{sc}$ as (Torrence \& Compo 1998):
\begin{equation}
\delta B_{f_{sc}}=\frac{\delta t^{1/2}}{C_{\delta} \psi_{0}}
\frac{Re(\tilde{B}(f_{sc}))}{s_{f_{sc}}^{1/2}}.
\end{equation}
Parameters used for reconstruction are  $C_{\delta}=0.776$,
$\psi_{0}=\pi^{-\frac{1}{4}}$, these are empirically derived for the
Morlet wavelet. Here, $s_{f_{sc}}$ (the order of time scale at
$f_{sc}$) is used to convert the wavelet transform $\tilde{B}$ to an
energy density (Torrence \& Compo 1998). At each frequency $f_{sc}$,
the reconstructed time series contain wave power within a frequency
window which is about 8.3\%$f_{sc}$ centered at $f_{sc}$.


The top panels of Fig. 3 show $\delta B_{x'}-\delta B_{y'}$
hodograph at four frequencies 0.96Hz, 0.74Hz, 0.52Hz and 0.20Hz
(from left to right) for P3 and P4 using data from C4. Results from
other three spacecraft are essentially the same. Each column
corresponds to results of one frequency. The magnitude of
$kv_A/\Omega_p$ determined by k-filtering is  1.35, 0.99, 0.75 and
0.39 for $f_{sc}=$0.96, 0.74, 0.52 and 0.2Hz, respectively.
 The bottom panels show
the $d\varphi/dt$ as a function of time, here $\varphi$ is the angle
that a magnetic field vector makes with the $\delta B_{x'}$ axis
such that \begin{equation} \varphi(t)=\arctan \left[ \frac{\delta
B_{y'}(t)}{\delta B_{x'}(t)} \right].
\end{equation}
Positive (negative) sign indicates that the polarization of the wave
is right- (left-) handed. It is clear from Fig. 3 that the
polarization of magnetic fluctuations in the plane perpendicular to
the wave vector ${\bf k}$ is dominantly right-handed. A closer look
of the bottom panels finds that the dominance of right-handed
polarization is more pronounced at 0.96Hz, 0.74Hz and 0.52Hz than at
0.2Hz. From Fig. 2e, we know that at 0.96Hz, 0.74Hz and 0.52Hz the
wave vectors ($kv_A/\Omega_p$ is 1.35, 0.98, and 0.75) are in the
dissipation range of the magnetic field power spectrum and at 0.2Hz
the wave vector ($kv_A/\Omega_p=0.39$) is at (or near) the spectral
break point where the spectrum switches from inertial range to the
dissipation range. This may suggest that the turbulence has
experienced some subtle change in the dissipation range where the
ion kinetic effect starts to kick in. For P1 and P2, the
polarization is also predominantly right-handed at all frequencies
studied by the k-filtering technique. The ${\bf B}_0$ in Fig. 3 is
the projection of the average magnetic field in the plane for the
whole period P3 (Figs. 3c and 3g) or P4 (Figs. 3d and 3h).


\section{Interpretations and discussions}

It is clear from Fig. 2b that the measured dispersion relation
cannot be explained by fast or Bernstein waves. These measured
dispersion relation points are quite scattered, and no dispersion
relationship of a single plasma wave can be uniquely identified from
the measured dispersion points, in accordance with the findings by
\citet{Narita2011}. From the k-filtering result, it is not clear if
we have observed kinetic Alfv\'en waves, convected coherent
structures, or a mixture of them (and others) in the solar wind. For
instance, while many of the data points may be interpreted to be on
the dispersion curves of the quasi-perpendicular propagating KAWs
within the uncertainties, they can equally be said to be on the
dispersion curve of convected coherent static structures within the
uncertainties.

From Fig. 3, except some less frequent anomalies one can see that
the major axis of magnetic ellipse is dominantly perpendicular to
${\bf B}_0$, and this has been interpreted as evidence of dominant
KAWs and not whistler waves (He et al. 2012), although He et al.
(2012) have to make assumptions on the wave propagation direction.
During periods P3-P4, the electron temperature is lower than the
proton temperature (for P1 and P2, PEACE $T_e$ data are not
available from the ESA Cluster Active Archive), kinetic slow waves
are not expected to exist due to strong damping. At
$80^\circ<\theta_{\bf kB}<90^\circ$, the dispersions of
fast/whistler waves are all similar to the red-solid line in Fig. 2c
and too far away from the observed dispersion points. Hence, the
wave polarization analysis in the plane perpendicular to ${\bf k}$
may support the interpretation that KAWs are an important turbulence
component at the ion kinetic scale turbulence.



However, KAWs interpretation has weakness. In the studied periods,
the polarization is dominantly right-handed ($\langle
d\varphi/dt\rangle$, the average $d\varphi/dt$, is positive) for
both $\theta_{\bf kB}>90^\circ$ and $\theta_{\bf kB}<90^\circ$.
According to Vlasov theory, in plasma of one ion species with
Maxwellian VDF, the magnetic field of KAWs (in the plasma frame) has
right-handed polarization when $\theta_{\bf kB}<90^\circ$ and
left-handed polarization when $\theta_{\bf kB}>90^\circ$ (here
$|\theta_{\bf kB}-90^\circ|<20^\circ$). The ion plasma betas in this
study are smaller than 1 (used by He et al. 2012) and $T_e$ is only
half of $T_i$. We find that the change of ion beta (0.6 - 1) and
$T_e/T_p$ (1 - 0.5) does not change the polarization of these waves.
One possibility of the observed $\langle d\varphi/dt\rangle$ at
$\theta_{\bf kB}>90^\circ$ is due to the large uncertainty of
$\theta_{\bf kB}$ from k-filtering, the $\theta_{\bf kB}$ of the
observed left-handed waves is actually smaller than 90$^\circ$.
Another weakness of KAWs is that the wave power along ${\bf k}$ is
found at least as strong as those in the direction parallel to ${\bf
B}_0$. This is shown in Fig. 4: at two frequencies $f_{sc}=$0.74 and
0.2 Hz, the reconstructed fluctuated magnetic field $\delta B_k$ and
$\delta B_\parallel$ along the direction of ${\bf k}$ and ${\bf
B}_0$ are shown as black and green lines within interval P3 (the
data are from spacecraft C4). At $f_{sc}=0.74$Hz, the fluctuated
magnetic field along the wave vector $\delta B_k$ is slightly
stronger than the fluctuated magnetic field along the background
magnetic field. At
 $f_{sc}=0.2$Hz, the fluctuated magnetic field along
the wave vector $\delta B_k$ is often twice as strong as $\delta
B_\parallel$. However, a kinetic Alfv\'en wave propagating along a
wave vector ${\bf k}$ is expected to generate no fluctuated magnetic
field this direction.

Since the measured dispersion points are quite scattered, they may
be seen as no clear dispersion (Narita et al. 2011), but the
superposition of different things such as waves and turbulent
structures.
 An alternative interpretation of the data is that
static small scale currents \citep{Perri2012} and 2D nonlinear
coherent structures (such as solitary monopolar and dipolar Alfv\'en
vortex filaments) with $k_\perp \gg k_\parallel$ populate at ion
kinetic scales. Monopolar Alfv\'en vortices are static structures
and dipole Alfv\'en vortices move with an arbitrary speed in the
plasma frame mainly in the direction perpendicular to ${\bf B}_0$.
The magnetic field fluctuations mainly occur in the direction
perpendicular to ${\bf B}_0$, which is the case shown in Fig. 3.
Indeed, dispersion relations of these static currents and structures
are flat in the solar wind frame (Fig. 2c).

To discuss the idea further, we conduct another
 \emph{polarization analysis in the
plane perpendicular to ${\bf B}_0$} and use ${\bf B}_0$ to replace
${\bf k}$ in Eq.(6) to construct new Cartesian coordinates. The
results for two frequencies $f_{sc}=0.74$Hz and 0.2Hz are shown in
Fig. 5. On average, the polarization of fluctuations ($\langle
d\varphi/dt\rangle$) can be either positive (Fig. 5b) or negative
(Fig. 5f). (The randomness of polarization in the plane
perpendicular to ${\bf B}_0$ is fine for KAWs since such a wave is
supposed to be linearly polarized in the plane and the presence of
many such waves can generate random overall polarization). At each
$f_{sc}$, the preference of polarization at one sense is weak
($|\langle d\varphi/dt\rangle|$ is smaller compared to those in the
plane perpendicular to ${\bf k}$). Magnetic fluctuations in Figs. 5c
and 5g consist of wave packets. The hodograms (Figs. 5d and 5h) of
the perpendicular field $\delta {\bf B}_\perp$ show that Cluster
went through regions of shear in the magnetic field (labeled as `C')
and rotations (for instance at 173.3s, 174.6s and 175.8s in Fig.
5d). Such coherent rotations are signatures of coherent Alfv\'en
vortices (Chemyrev et al. 1988, Volwerk et al. 1996). The rotational
sense changes frequently (blue and red denote opposite rotations).
When the polarization changes (color changes between red and blue)
in Figs. 5d and 5h, the $\delta B_\perp$ does not experience any
appreciable change in either magnitude or direction: these
polarization changes do not correspond to discontinuities or
currents sheets.

Alfv\'en vortices (drift Alfv\'en vortices) are 2D tubular
structures and exist in homogeneous (inhomogeneous) plasmas. The
observed polarization depends on the trajectory of satellites across
the monopolar or dipolar vortex: it can be elliptical (linear or
circular), right- or left-handed as a function of the trajectory. In
homogeneous plasmas, there is no limit to the dimension (radius) of
Alfv\'en vortices.In inhomogeneous plasmas, the theory of Alfv\'en
vortices valid for scale sizes of ion inertial length and ion Larmor
radius can be found in Chmyrev et al. (1988) and Onishchenko et al.
(2008), respectively. The dimension of measured Alfv\'en vortices
tends to be the order of the ion gyroradius (Sundkvist et al. 2005).
Such drift Alfv\'en vortices are generated naturally in plasmas with
strong gradients when the drift velocity of particles $V_d=-\nabla p
\times {\bf B}/neB^2$ is comparable to their thermal velocity
(Petviashvili \& Pokhotelov 1992), or equivalently the density scale
size matches the ion Larmor radius (Sundkvist \& Bale 2008).

In the solar wind at 1AU the ion drift velocity is small due to weak
inhomogeneity and Alfv\'en vortices may be used to describe these
rotational structures. In Fig. 5c (0.74Hz), a wave packet typically
lasts $\Delta t=$6-8s. The dimension of such wave packet in the
direction perpendicular to ${\bf B}_0$ is $V_f\Delta
t/\sin(\theta_{VB})= 3400\sim 5300$km, suggesting that the radius of
such structures is $a=13\sim 20.5\rho_i$. Similar structures with
discontinuities have been studied in the context of the solar wind
(Verkhoglyadova et al. 2003), and have been found in the Earth's
magnetosheath (Alexandrova et al. 2006) and Saturn's magnetosheath
(Alexandrova \& Saur 2008). The wave number determined by the
k-filtering is $kv_A/\Omega_p=0.98$ at 0.74Hz, corresponding to a
scale of 1160km, in accordance with an Alfv\'en vortex with
$a=20.5\rho_i$ (13.5$\rho_i$) if the vortex boundary corresponding
 to the third (second) zero of Bessel function of the
first kind. Similarly at 0.2Hz (Fig. 5g), a wave packet typically
lasts 20-22s. The dimension of such a wave packet in the direction
perpendicular to ${\bf B}_0$ is $V_f\Delta t/\sin(\theta_{VB})=
13200\sim 14500$km. The radius of such a structure is $a=96 \sim 106
\rho_i$, considerably larger than that at higher frequency 0.74Hz as
we would expect.


The theory of solitary Alfv\'en vortex is based on single-fluid MHD
and assumes incompressibility (Petviashvili \& Pokhotelov 1992). At
the scales we studied, it is clear from Fig. 3 that $\delta
B_\parallel \ll \delta B_\perp $ so the turbulence incompressibility
is approximately met.
 The limitation of solitary Alfv\'en vortex interpretation is
the difficulty to explain the polarization in the plane
perpendicular to ${\bf k}$ in Fig. 3 with solitary Alfv\'en
vortices. This is because that the theory of solitary Alfv\'en
vortex assumes $\delta B_\parallel \approx 0$. A nonzero $\delta
B_\parallel$ is necessary to explain the dominantly right-handed
polarization in the plane perpendicular to ${\bf k}$ if the wave
vector is perfectly perpendicular to ${\bf B}_0$. The k-filtering
analysis finds that the wave vector mainly points to directions
nearly perpendicular to ${\bf B}_0$. One would expect that the wave
polarization in the plane perpendicular to ${\bf k}$ does not have a
preference in either left-handed or right-handed sense when a
spacecraft passes through many of such structures.
 A theory of
solitary Alfv\'en vortex including small compressibility is needed.

\section{Summary and conclusion}

In summary, the application of the k-filtering technique and wave
polarization analysis to turbulence at the proton gyroscales in the
fast solar wind found the following: Turbulence at these scales
slowly (compared to the Alfv\'en speed) propagates in the directions
nearly perpendicular to ${\bf B}_0$.
The fluctuated magnetic field
 in the frequency range 0.07-1.1Hz shows higher $\delta B_\perp$ than
$\delta B_\parallel $ and has dominantly right-handed polarization
in the plane perpendicular to the wave propagation direction at both
$\theta_{\bf kB}<90^\circ$ and $\theta_{\bf kB}>90^\circ$. The
polarization of the fluctuations is elliptical with a regular change
of polarization from right to left-handed. Wave polarization is
quite random in the plane perpendicular to the background magnetic
field and is consistent with the interpretation of Alfv\'en
vortices. The wave polarization in the plane perpendicular to wave
vector ${\bf k}$ is more consistent with linear kinetic Alfv\'en
waves than Alfv\'en vortices. It is found that no dispersion
relation of a single plasma wave mode can be uniquely identified
from the measured wave/turbulence dispersion plots.

We have discussed the pros and cons  of KAWs and coherent structures
in the interpretation of the solar wind turbulence at ion kinetic
scales. A plausible scenario is that at such scales KAWs and
coherent structures co-exist in the fast solar wind described in
this study. It is noted that further validation of the k-filtering
technique may be needed when the analyzed signal contains a mixture
of coherent structures and plane waves with random phases, not just
plane waves with random phases alone. On the other hand, one may see
certain similarity between a series of intermittent coherent
structures of similar sizes passing a spacecraft and a plane wave
with a random phase passing a spacecraft. We plan to publish such a
validation elsewhere.





\begin{acknowledgments} All Cluster data are obtained from the ESA
Cluster Active Archive. We thank the FGM, CIS and PEACE instrument
teams and the ESA Cluster Active Archive. This study is supported in
part by the National Natural Science Foundation of China (40904047
and 41174154). XL thanks helpful discussions with Olga Alexandrova
and Elizabeth Lucek.
\end{acknowledgments}


\clearpage

\begin{figure}
\includegraphics{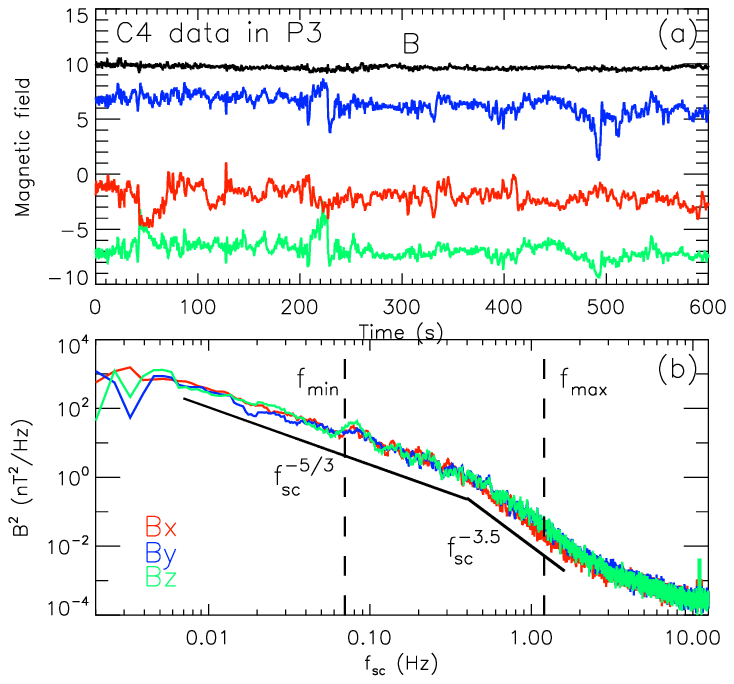}
\caption{Spectra of typical magnetic field components $B_x$, $B_y$
and $B_z$ measured by FGM from  04:10UT to 04:35UT on 29/02/2004.
The vertical dashed lines denote the frequency range that
k-filtering technique is applied. The spectral flattening above
2.4Hz is due to the FGM reaching the noise floor.\label{fig1}}
\end{figure}

\begin{figure}
\includegraphics[width=15cm]{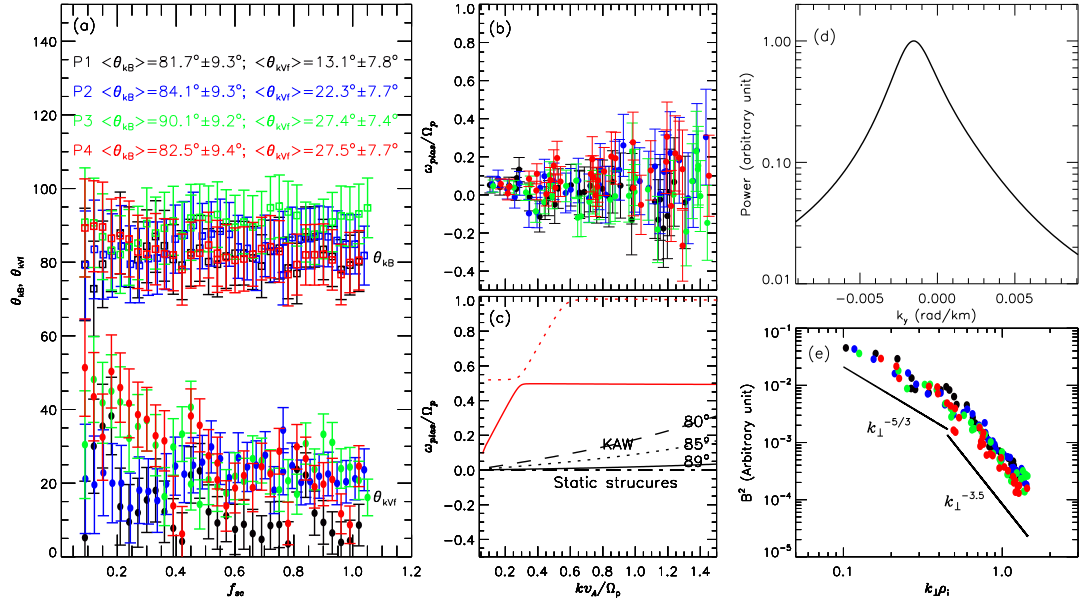}
\caption{(a) Angles $\theta_{\bf kB}$ (squares) and $\theta_{{\bf
kV}_f}$ (solid dots) with related uncertainties computed by using
the k-filtering technique during four time intervals. (b) Measured
wave dispersion (filled dots), with estimated error bars. (c) The
dispersion relation curves, computed from linear Vlasov theory,
represent waves propagating at several observed angles $\theta_{\bf
kB}$: black lines are kinetic Alfv\'en waves propagating at
80$^\circ$ (dashed), 85$^\circ$ (dotted) and 89$^\circ$ (solid); the
remaining red (solid and dotted) curves represent fast and Bernstein
waves propagating at 89.5$^\circ$. The proton angular gyrofrequency
is $\Omega_p=2\pi f_{ci}$.
The dot-dashed line represent static structures. 
(d)Power spectral density of magnetic field fluctuations as a
function of $k_y$ at spacecraft-frame frequency 0.51Hz in P3. (e)
The magnetic field $k_\perp$ spectra of all the four measured time
intervals.
 }
\end{figure}


\begin{figure}
\includegraphics[width=22cm]{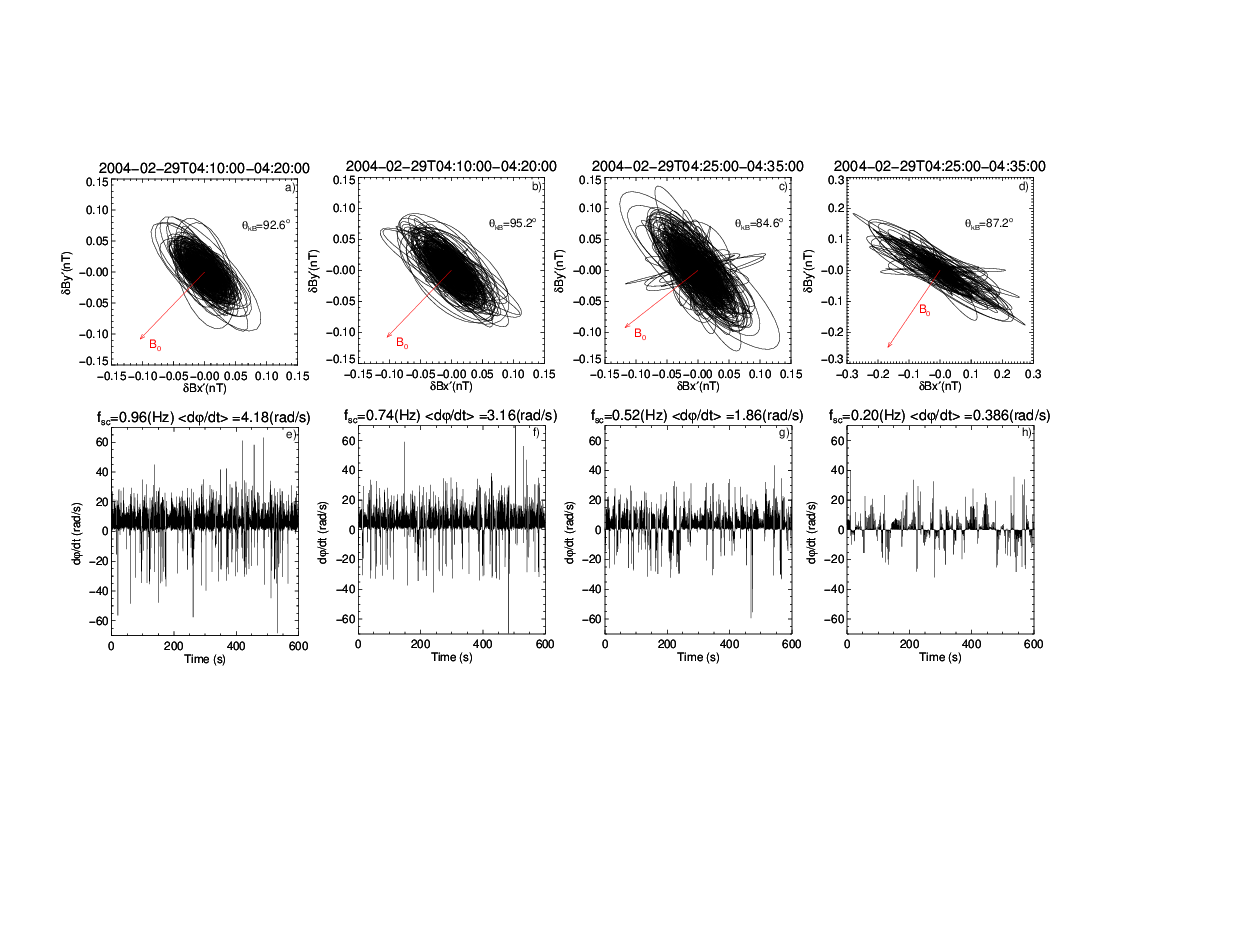}
\caption{Polarization analysis in the plane perpendicular to ${\bf
k}$. Top panels: $\delta B_{x'}-\delta B_{y'}$ hodograph at
frequencies (a) $f_{sc}=0.96$Hz, (b)0.74Hz in P3,  and (c) 0.52Hz,
(d) 0.2Hz in P4. The bottom panels display the corresponding
polarizations at these frequencies. The wave propagation angle
$\theta_{\bf kB}$ is a) 92.6$^\circ$, b) 95.2$^\circ$, c)
84.6$^\circ$, and d) 87.2$^\circ$. The mean values of $\langle
d\varphi/dt\rangle$ are all positive. The wave frequencies in the
solar wind frame determined by
the k-filtering are all positive. 
}
\end{figure}

\begin{figure}
\includegraphics{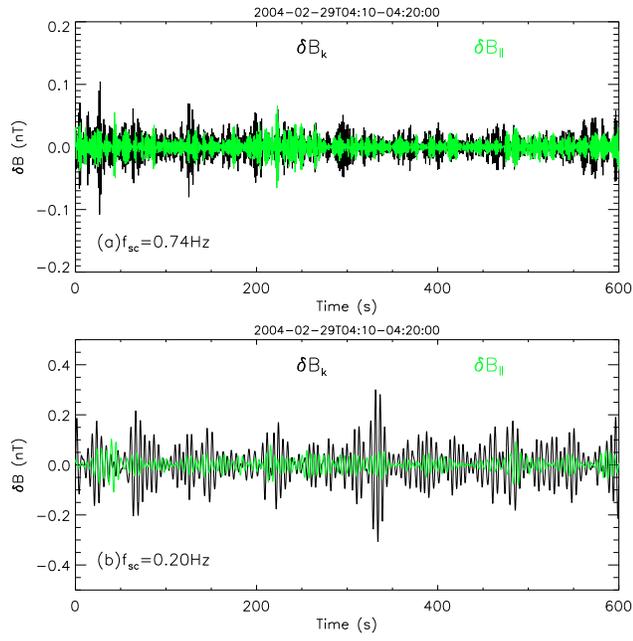}
\caption{Reconstructed time series of fluctuated magnetic field
along wave vectors found by the k-filtering technique and along the
background magnetic field ${\bf B}_0$ for the interval P3 at two
spacecraft frequencies: (a) $f_{sc}=0.74$Hz, and (b) $f_{sc}=0.2$Hz.
The data are taken from spacecraft C4. }
\end{figure}

\begin{figure}
\includegraphics[width=16cm]{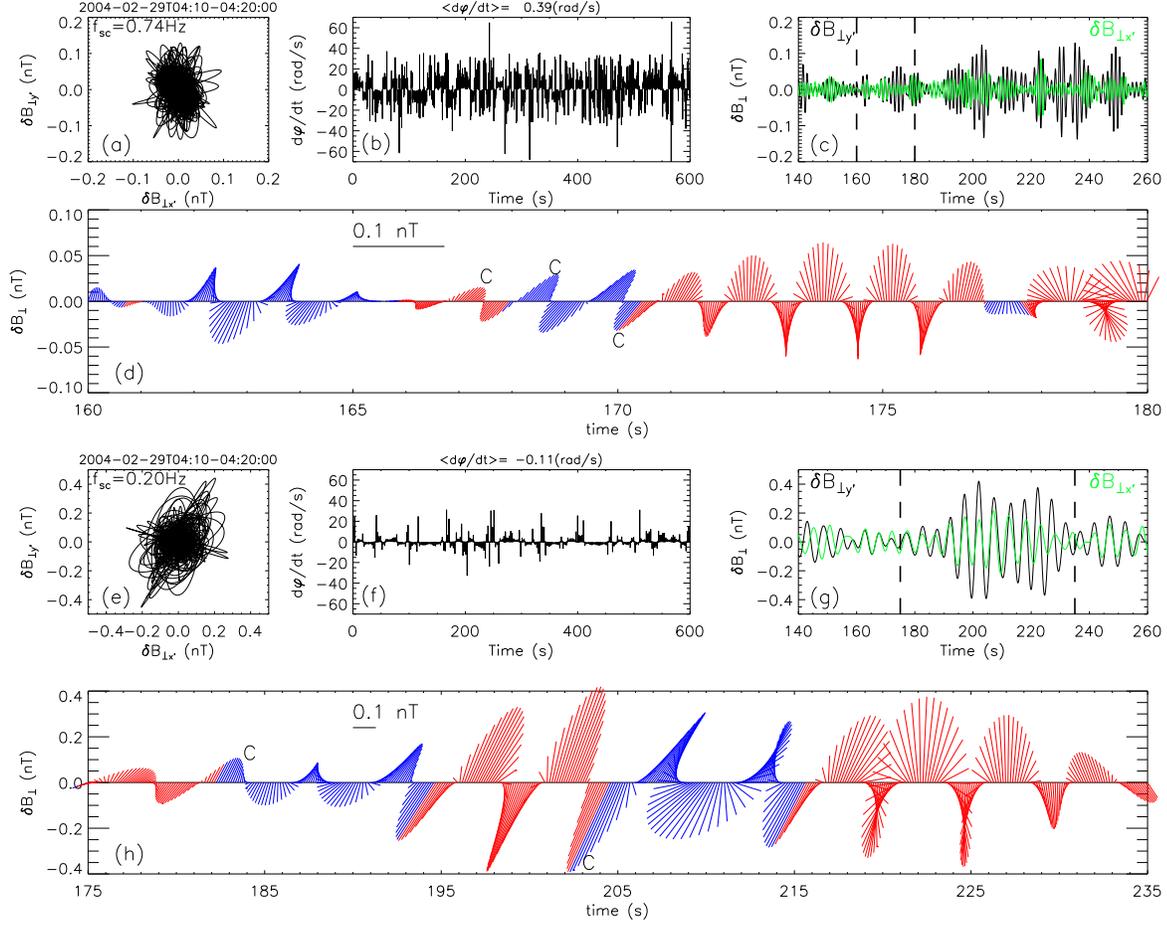}
\caption{Polarization analysis in the plane perpendicular to ${\bf
B}_0$ at $f_{sc}=0.72$Hz (a,b,c,d) and 0.21Hz (e,f,g,h) for interval
P3. (a) and (e): $\delta B_{\perp x'}-\delta B_{\perp y'}$
hodograph, (b) and (f): $d\varphi /dt$, (c) and (g): representative
waveforms, magnetic field hodograms for the region between the two
vertical lines are shown in (d) and (h). Blue and red denote right-
and left-handed polarization. At the two frequencies, the wave
vectors determined by k-filtering are $kv_A/\Omega_0$ =0.981 and
0.354. The data are taken from spacecraft C4.
 }
\end{figure}


\end{document}